\documentclass[lettersize,journal]{IEEEtran}
\usepackage{amsmath,amsfonts}
\usepackage{orcidlink}

\usepackage{array}
\usepackage[caption=false,font=normalsize,labelfont=sf,textfont=sf]{subfig}
\usepackage{textcomp}
\usepackage{stfloats}
\usepackage{url}
\usepackage{verbatim}
\usepackage{graphicx}
\usepackage{cite}
\usepackage{scalefnt}
\usepackage{pgfplots}
\usepackage{multirow}
\pgfplotsset{compat=1.11}
\usepackage{svg}
\usepackage[ruled]{algorithm2e}
\usepackage{algpseudocode}
\DeclareUnicodeCharacter{2217}{*} 

\begin{document}

\title{Enhancing Semantic Document Retrieval: Employing Group Steiner Tree Algorithm with Domain Knowledge Enrichment }

\author{{Apurva Kulkarni \orcidlink{0000-0002-9215-2049},  
\and Chandrashekar Ramanathan \orcidlink{0000-0002-3330-8365}, and \and
 Vinu E Venugopal \orcidlink{0000-0003-4429-9932}}
\thanks{This work is supported by Karnataka Innovation \& Technology Society, Dept. of IT, BT and S\&T, Govt. of Karnataka, India}}



\maketitle
\begin{abstract}
Retrieving pertinent documents from various data sources with diverse characteristics poses a significant challenge for Document Retrieval Systems. The complexity of this challenge is further compounded when accounting for the semantic relationship between data and domain knowledge. While existing retrieval systems using semantics (usually represented as Knowledge Graphs created from open-access resources and generic domain knowledge) hold promise in delivering relevant outcomes, their precision may be compromised due to the absence of domain-specific information and reliance on outdated knowledge sources. In this research, the primary focus is on two key contributions: a) the development of a versatile algorithm- `Semantic-based Concept Retrieval using Group Steiner Tree' that incorporates domain information to enhance semantic-aware knowledge representation and data access, and b) the practical implementation of the proposed algorithm within a document retrieval system using real-world data. The proposed algorithm leverages domain information to construct knowledge graphs and utilizes semantic grouping through the Group Steiner Tree technique to identify relevant concepts for accessing pertinent data. To enable semantic-based document retrieval, the implementation integrates the proposed algorithm into the SemDR (Semantic-based Document Retrieval) system, which facilitates the integration and querying of heterogeneous data while considering semantic relationships. To assess the effectiveness of the SemDR system, research work conducts performance evaluations using a benchmark consisting of 170 real-world search queries. Rigorous evaluation and verification by domain experts are conducted to ensure the validity and accuracy of the results. The experimental findings demonstrate substantial advancements when compared to the baseline systems, with precision and accuracy achieving levels of 90\% and 82\% respectively, signifying promising improvements.
\end{abstract}

\section{Introduction}
 The Document Retrieval (DR) involves searching for and obtaining specific documents from a collection of heterogeneous data sources using a search term or query~\cite{i1}.  The selection of a retrieval mechanism is influenced by various factors, including the nature of documents, user needs, and available resources. Traditional approaches rely on techniques such as natural language processing, machine learning, statistical models, neural networks, and semantic-based models to retrieve relevant information. Keyword-based models rely on accurate matches between a user's query and specific keywords or phrases in a document~\cite{k0,k1,k2,k3,k4}, while concept-based models~\cite{c1,c2,c3} and semantic models~\cite{s1,s2,s3,s4} aim to understand the underlying concepts behind a user's query to give more accurate and comprehensive search results. Models used for retrieval include the Boolean model, Vector Space Model (VSM), probabilistic models, and language models pre-trained on extensive text data~\cite{bool1,bool2,vec1,vec2,ml1,prob2}. Different models are often leveraged in combination to improve the accuracy and relevance of search results (In Section VI, a comprehensive analysis of the existing literature is presented). The motivation for this work stems from the challenges and limitations observed in traditional document retrieval approaches.
 \subsection{Motivation}
Let's consider a scenario where a user wants to retrieve documents related to their search query `Fiber Export'. In this case, it is understood that the documents in the agriculture field are relevant to the search. While documents containing the words `Fiber' or `Export' would be an obvious choice, the ideal retrieval would focus on any type of fiber material being exported. The user's main interest lies in obtaining files that capture export details of any fiber material.
 \begin{figure}[!h]
    \centering
      \includegraphics[width=87.9mm,height=45mm]{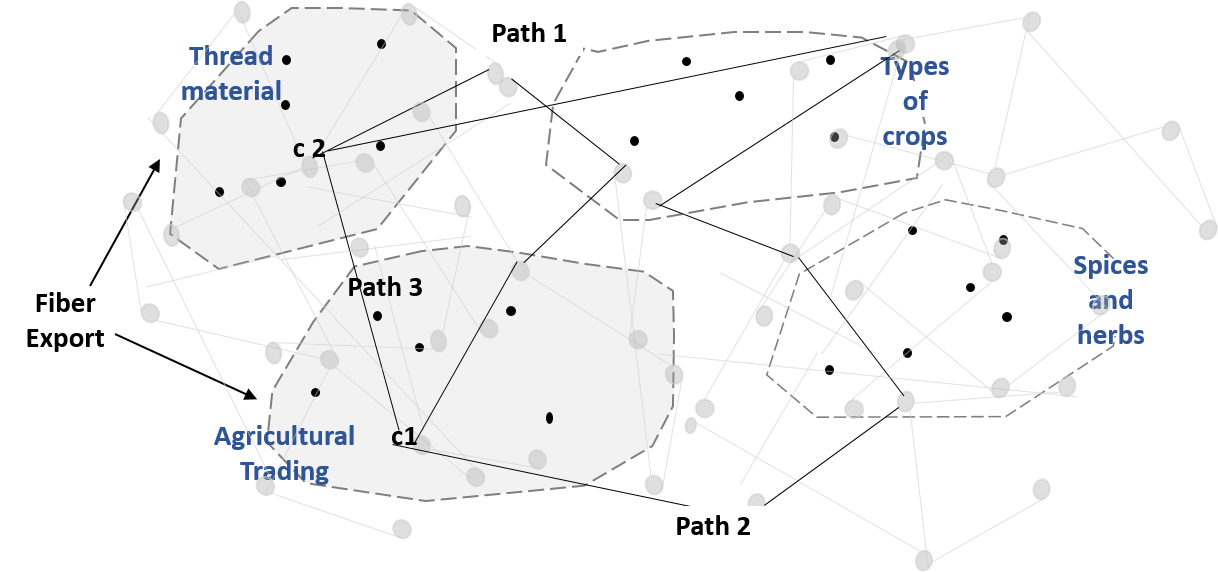}
      \caption{Motivational Example: Identifying relevant data points in agriculture domain for search string}
      \label{f111}
    \end{figure}
To further illustrate this example, we can refer to the graph depicted in Figure~\ref{f111}. This graph represents important entities --in Section II, we define them as `concepts'-- and their relationships within the agriculture domain. The graphical structure serves as a representation of domain-specific knowledge, showcasing the significance of semantics. The relevant concepts in the graph are denoted by dotted lines, indicating their membership in different groups. Identifying the concepts associated with `Fiber' and `Export' in the graph becomes more intuitive by considering the concepts belonging to Group `Thread Material' and Group `Agriculture Trading'. Several paths link concepts from these two groups, and among them, Path 3 stands out as it directly connects Group `Thread Material' and Group `Agriculture Trading,' indicating concepts with a high semantic correlation to the search words.
However, it is worth noting that Path 1 and Path 2, while connecting concepts from different groups (`Types of crops' and `Agricultural Trading'), include concepts that are not directly relevant to the search words.
This example highlights the value of leveraging domain knowledge for semantic grouping and graph traversal, resulting in more effective retrieval of semantically aware results.

 In the proposed research, we introduce a new method that utilizes a Concept Graph to represent domain knowledge\cite{cikm}. This graph captures entities, and their interactions, presenting a comprehensive view of the domain. By leveraging the relationships and connections within the graph, our retrieval system can accurately identify and retrieve contextually relevant information that aligns with the user's intent. 

\subsection{Research Contribution} The key characteristics of our work are:
\begin{itemize}
    \item The primary contribution of the research work is the novel algorithm- `Semantic-based Concept Retrieval using Group Steiner Tree’ that uses concepts and semantics.
    \item The proposed algorithm has a generic nature and can be extended to various applications, including entity resolution, information retrieval, text summarization, question-answering systems, compensation systems, and ontology alignment. In this paper, we specifically focus on expanding its application to document retrieval.
    \item The SemDR (Semantic-based Document Retrieval) system  preserves heterogeneous data sources in their original forms while enabling context-relevant search. The proposed system is implemented and evaluated on real-world data\footnote{Data is received from the Government of Karnataka}. The search queries are based on the needs of domain users and the result of these search queries are evaluated by domain experts and various users.
\end{itemize}
\subsection{Organization of the work} Continuing with the research narrative presented in Section I, Section II discusses the construction of a semantic concept graph that builds the foundation for the proposed algorithm discussed in Section III. Section IV delves into the implementation of the proposed algorithm, presenting both a logical design of the proposed SEMDR system  and a case study that supports its implementation. Additional details on testing, benchmarking, and the findings of the study are provided in Section V. Section VI highlights a comprehensive analysis of related literature. Finally, the concluding segment of the paper summarizes the significant contributions made by the research and outlines potential pathways for future research and exploration.

\section{Semantic Concept Graph}
To enhance document retrieval through semantic understanding, it is vital to grasp both the contextual aspects of the domain and the interconnections and arrangement of knowledge. Employing a graph-based structure to capture domain knowledge enables the organization of information in a structured manner and facilitates navigation through traversal. Our proposed method employs a graph-based approach for the document retrieval problem. We have added a semantic component to the graph, which establishes connections between concepts, resulting in the creation of a \emph{semantic concept graph} that includes a semantic layer to identify concepts relevant for the retrieval. The captured semantics aid in retrieving pertinent documents.

 \subsection{Building Semantic Concept Graph}
\begin{figure}[!h]
    \centering
      \includegraphics[width=88.9mm,height=55mm]{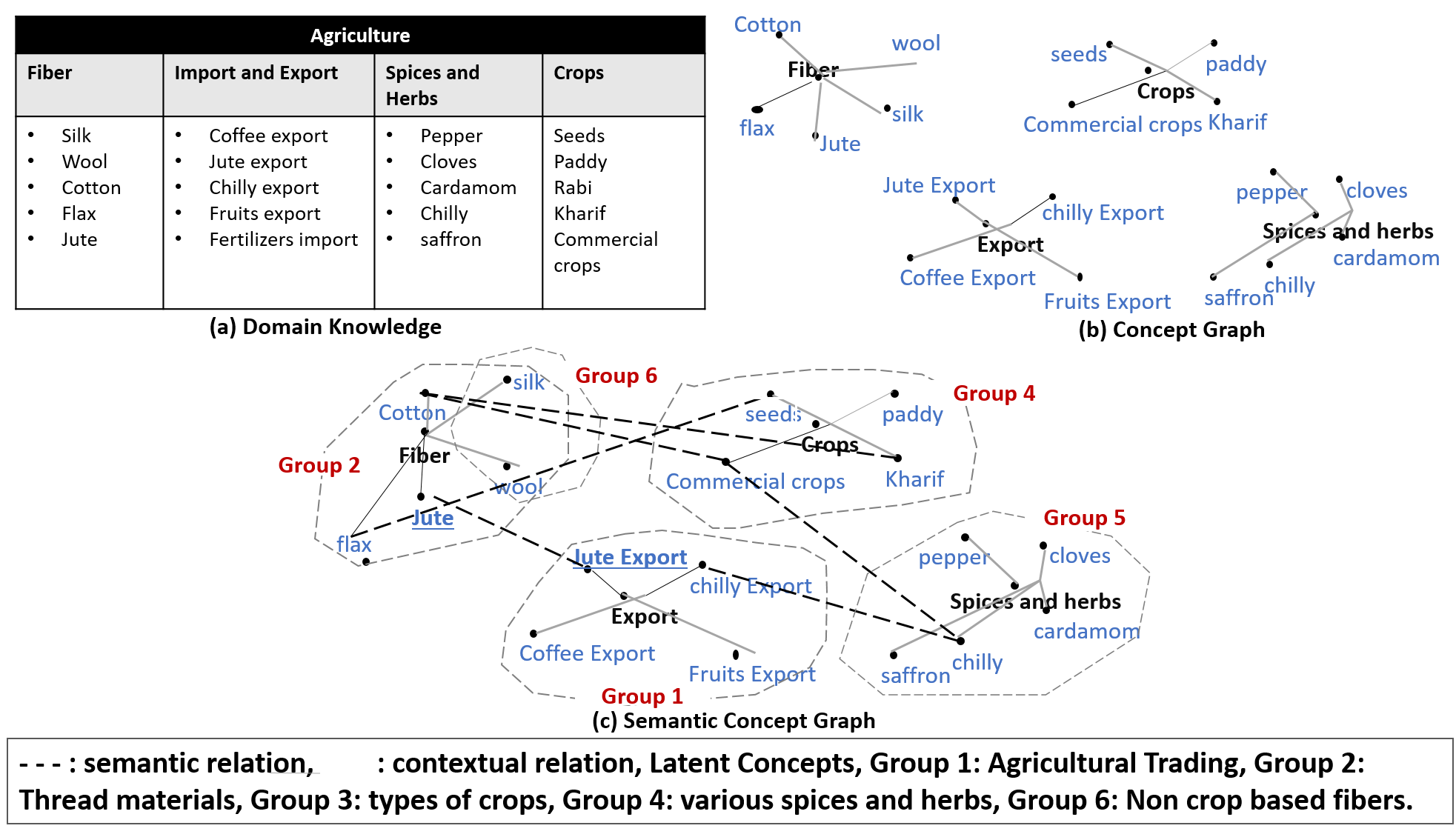}
      \caption{An example illustrating the creation of a semantic content graph for  agriculture domain}
      \label{f21}
    \end{figure}

One effective way to represent the information is by building a knowledge graph, comprising nodes representing entities and edges representing relationships~\cite{p1}, that provides a comprehensive representation of real-world entities and their attributes. Figure~\ref{f21} sub-figure (a) illustrates the agriculture domain concepts, while sub-figure (b) provides a graphical representation of these concepts. The proposed document retrieval (DR) approach primarily focuses on identifying relevant concepts (both explicit and latent) within the domain context and their interconnections. 

The explicit concepts derived from domain information are represented as nodes, and the hierarchical relationships (subclass or superclass) are known as contextual relations. 
We have considered a \emph{semantic proximity} metric (defined later), which decides the degree of similarity or relatedness between concepts based on their meanings or semantic representations, to further reinforce the concept graph. Using this measure, concepts with similar properties are linked together, forming semantic relations. Additionally, concepts sharing common properties are grouped into latent concepts based on their commonalities; represented in red bold font in the Figure~\ref{f21} sub-figure (c).\\
\textbf{Semantic Concept Graph.} 
 A semantic concept graph is an edge-labelled graph $G =(C, R, W)$  where $C$ corresponds to a finite set of concepts,  $R$ denotes a finite set of undirected edges between two concepts, and $W$ indicates weights associated with edges. The set of concepts $C$ is a collection of \textit{direct concepts} and \textit{latent concepts}. That is, $C= C_{direct} \cup C_{latent}$, where $C_{direct}$ is a concept derived from domain knowledge and $C_{latent}$ indicates a logical node that refers to a group of similar concepts. The set of relations $R$ is a collection of\textit{ contextual relations} and \textit{semantic relations}. That is, $R =R_{contextual} \cup R_{semantic}$, where Contextual relation $R_{contextual}$ captures the hierarchical connections and dependencies among various concepts in the domain and semantic relation $R_{semantic}$ links the concepts with high semantic proximity. In this work, we determined the semantic proximity of two concepts using the Wu-Palmer Similarity score~\cite{wu}. The semantic score varies between 0 to 1, where value 1 denotes the maximum similarity. The default threshold is set as 0.9 to achieve refined results.
  \par Figure~\ref{f21} (c) depicts the agriculture domain knowledge, including concepts such as cotton, silk, crops, and kharif. The blue-highlighted concepts are sub-concepts of the concepts denoted by bold black letters. In the figure, 
  concepts are semantically grouped to define latent concepts. Each group represents a latent concept. For example, Group 1 represents the latent concept ``agricultural transport", encompassing concepts related to transporting agricultural goods. Group 2 connects various thread materials, including both plant-based and animal-based threads like silk and wool. Furthermore, Group 6, a subgroup within Group 2, categorizes specific concepts related to animal-based thread materials. The graph also incorporates additional latent concepts, which capture meaningful associations and relationships between concepts, enriching the semantic meaning and understanding of the graph.

   \subsection{Assigning Weights to Semantic Concept Graph} After the construction of the semantic graph, we proceed to determine the weights assigned to each node and edge within the graph. These weight calculations can vary depending on the specific application. In our case, which involves document retrieval, we utilize two weight calculation functions, namely \textit{concept\_domain} and \textit{relation\_score} that are based on the content of the documents mapped to the concepts. Both these functions are defined later in Section IV. By leveraging the weights assigned to the graph's edges, our DR algorithm will navigate the graph and retrieve relevant concepts based on their semantic similarities and relationships.

    
\section{Proposed Algorithm: Semantic-based Concept Retrieval using Group Steiner Tree}
To identify the documents relevant to the search word, a crucial intermediate step is to determine the concepts that are most relevant to the search word. 
We propose Algorithm~\ref{alg1} to 
identify the concepts that closely align with the search word from the weighted semantic concept graph. The algorithm operates through the following steps:
    \begin{itemize}
    \item \textit{Identifying the anchor concepts:} Determine the ``anchor concepts'' in the graph by identifying concept nodes that exhibit a semantic proximity score greater than 0.9\footnote{In this particular implementation set up, we established a threshold of 0.9 after experimenting with different values and taking into account the insights of domain experts.  } with the  words in the search string.
    \item \textit{Identifying the latent concepts:} By considering the Concept Graph containing \emph{Semantic Groups} and \emph{Anchor Concepts}. \\\textit{Note: The task involves relaxing the Anchor concepts to encompass the respective group of concepts (latent concept) having anchor concept.}
    \item \textit{Finding the relevant concepts.}  The key idea for identifying relevant concepts is that these nodes should be tightly connected to many anchor concepts. To formalize this intuition, we consider three factors: (i) the relevant documents lie on the intersection or union of anchor concepts, (ii) short paths are preferred, and (iii) paths with higher weights are better. These criteria are captured by the notion of a Steiner Tree.
    \end{itemize}
     \par Given an undirected and weighted concept graph $(C, R, W)$ with nodes $C$, edges $R$, and weights $\{w_{ij} \in W | w_{ij}\ge 0\}$ for the edge $(i,j)$ between nodes $i$ and $j$ and given a set $T \subseteq C$ of nodes called terminal nodes or anchor concepts, compute a tree $(C^{*}, R^{*}, W^{*})$ where $R^{*}\subseteq R, C^{*} \subseteq C$ that connects all terminals and has minimum cost in terms of total edge $∗$ weights, i.e., $Min(\sum_{(i,j)\in R^{*}} w_{ij})$ 
     \par In the case of two terminals, the solution can be obtained by finding the shortest path that connects two or more terminal nodes. However, our use case involves multiple terminals referred to as anchors. Additionally, these terminals are grouped into sets, where each set corresponds to a search keyword. In this scenario, it is sufficient to include at least one terminal from each set in the Steiner Tree. This problem, which encompasses these additional complexities, is commonly referred to as computing a Group Steiner Tree (GST)~\cite{gst1}.
    \par Given an undirected and weighted concept graph $(C, R, W)$ and given groups of terminal nodes $T$, 
    for each $A\in T$ where $A\subseteq R$, compute the minimum-cost tree $(C^{*} , R^{*})$ that connects at least one node from each $A \in T$ with minimum weighted path.
    
\begin{table}[!ht]
    \centering
    \caption{Test-cases : The table elaborates on various use cases and the approaches used by the proposed algorithm}
    \begin{tabular}{|p{0.02\linewidth}|p{0.1\linewidth}|p{0.75\linewidth}|}
    \hline
\textbf{No.} &\textbf{Use-case} &\textbf{Relevant Concepts Identified by Proposed Algorithm} \\
\hline
1.&No anchor concept identified & No relevant concepts retrieved. In Figure~\ref{f21} subfigure (C), let's consider a different search string, namely `Silk Import'. It is observed that  only one concept is identified as an Anchor concept-`Silk', which is most dominantly related to the search string with high semantic proximity  score.
Subsequently, files associated with the identified Anchor concept-`Silk', are retrieved from the data repository.\\
\hline
2.&Single anchor concept identified & Documents related to single concept retrieved. In the context of Figure~\ref{f21} subfigure (C), let's examine the search string `Education Facilities'. Upon analysis, it is determined that this search string lacks any meaningful semantic relativeness with the domain or the existing Anchor concepts.
As a result, the system fails to identify any Anchor concept corresponding to the search string `Education Facilities'. Consequently, there are no relevant Anchor concepts available to guide the retrieval process.\\
\hline
3.&Multiple anchor concepts are identified which are directly linked & The files associated with these identified Anchor concepts are fetched, and the files common among all the concepts are given the most importance. In Figure~\ref{f21} subfigure (c), we illustrate a scenario involving a search string, denoted as `Chilly Export'. This search string is mapped to two Anchor concepts: Chilly' and `Chilly Export'. It is important to note that a direct path exists between these two Anchor concepts.
To accomplish the retrieval process, the files associated with these identified Anchor concepts are fetched from the data repository. These retrieved files are then subjected to a comparison process. Specifically, the files common to both Chilly' and Chilly Export' are isolated for further analysis or consideration. \\
\hline
4.&Multiple anchor concepts are identified which are indirectly linked & The proposed algorithm is employed to identify the optimal path. Files associated with concepts retrieved in the optimal path are fetched, and the files common among all the concepts are given the most importance. In Figure~\ref{f21} subfigure (c), let's consider the search string `Fiber Export' as an example. The search process identifies two Anchor Concepts: Fiber' and `Agriculture Export'. These Anchor Concepts are connected by multiple paths, giving rise to various potential routes between them In this scenario, path Fiber, Jute, Jute Export, and `Agriculture Export are chosen and these concepts are further used by the retrieval process to fetch documents.\\
\hline
\end{tabular}
\label{usecases}
\end{table}     
Technically, the objective is to discover a Group Steiner Tree (GST) that encompasses latent concepts, linking a maximal count of Anchor concepts while minimizing the associated expenses. The pertinent concepts that aid in connecting the anchor concepts are the concepts of interest because they are semantically linking the anchor documents which aid in fetching the relevant documents. The use cases discussed in Table~\ref{usecases} elaborate on the diverse considerations employed by the algorithm during the GST identification process.

\subsection{Proposed Algorithm} 
Since concept graphs can easily grow vast and complex, volume is one of the main issues to be addressed. As the amount of data grows, the performance may suffer, creating an unsatisfactory user experience and long response times. Traversing the entire graph for search interest would be unfeasible. To make the graph efficiently queryable, it is necessary to use traversal algorithms to facilitate seamless graph processing.

    \begin{algorithm}
    \caption{Generic Framework}
    \label{alg1}
    \small{
    \KwData{Domain Knowledge \(Triples_{S-P-O}\), Search Query \(Data_{SQ}\)}
    \KwResult{Relevant Concepts \(Result_{RC}\)}
    $v1\gets Triples_{S-P-O}$\\
    $v2\gets generate\_semantic\_concept\_graph(v1)$\\
    $G\gets Intialize\_weights\_(v2) //G=(C, R, W)$\\
    $T\gets Terminal\_Nodes(G,Data_{SQ})$\\
    $Result_{RC}\gets$$Semantic\_Concept\_Retrieval\_GST(G,T)$\\
    \textbf{Return} Relevant Concepts \(Result_{RC}\)}
    \end{algorithm}
The suggested method centers around easing traversal by taking concept groups (latent concepts) into consideration. Initially, the algorithm identifies anchor concepts by assessing their semantic proximity, requiring a value greater than 0.9. The algorithm's objective is to determine the potential concepts that establish links between these anchor concepts. The intuition is that these potential concepts, referred to as relevant concepts, could serve as common semantic points, sharing similar semantics among the interconnected anchor concepts. Consequently, these relevant concepts would be an ideal choice for identifying the relevant documents.
\par In order to identify the relevant concepts, an effective approach could involve combining the minimum spanning tree and the shortest path algorithms to establish an optimal path connecting the anchor concepts. The Steiner Tree problem, which is an optimization problem, is utilized to find the tree with the lowest cost that spans a specific set of terminal nodes (the anchor concepts). This tree includes all the terminal nodes and the edges that connect each terminal node to at least one other terminal node, and it may also incorporate additional nodes that contribute to connecting the terminal nodes. It is important to note that these additional nodes, which connect the terminal nodes, represent the potentially relevant concepts that we are seeking to identify.
\par Solving the Steiner Tree problem involves the algorithm exploring all conceivable combinations of edges and nodes to determine the optimal tree. However, as the problem size grows, the computational complexity becomes impractical due to the exponentially expanding solution space. Consequently, the Steiner Tree problem is categorized as NP-hard, implying the absence of any known polynomial-time algorithm capable of solving it exactly for all instances. As a result, achieving an optimal solution usually necessitates utilizing approximation algorithms or heuristics that yield solutions of good quality, although not guaranteed to be optimal, within a reasonable timeframe.
\par In this work we propose a heuristic-based method, with the Greedy algorithm. The Group Steiner Tree (GST) problem extends the traditional Steiner Tree problem by aiming to find a tree with the lowest cost that spans a group of concepts, referred to as groups or latent concepts. This generalization allows for the consideration of common connections among different groups, thereby minimizing the overall cost of the network. We define the Group Steiner Tree in context as follows:

Given a weighted semantic graph $G=(C, R, W)$, a set of terminal nodes $T$  ($T\subset C$) where a collection of latent concepts (or semantic groups) $L$ are defined as: 

$T = \{ c | c \in C \cap (\exists word \in Tokenized(Search\_String) \cap$  $semantic\_score(word,c)>0.9) \}$.

 $L = \{ S | S \subset C \cap (\exists d \in S | \forall c \in S \symbol{92}\{d\},semantic\_score(c,\\d)>0.9)\}$

\par The goal is to identify a GST that spans across Semantic Groups, connecting a maximum number of Anchor concepts while minimizing the associated cost. The Anchor concepts, along with the additional concepts that connect them, are considered the Relevant Concepts of interest.

\begin{algorithm}
\caption{Semantic Concept Retrieval by GST}\label{alg2}
\vspace{1mm}
\small{
\textbf{Input:}
\\Weighted Semantic Concept Graph $G=(C, R, W)$,\\ 
Terminal Nodes $T\gets \{T_{1},...,T_{n}\}$\\
Latent Concepts $L\gets \{L_1, ..., L_n\}$\\
(referred to as Terminal Nodes)\\
\textbf{Output:}\\Relevant Concepts \(Relevant\_Concepts\)\\
\textbf{Initialize:}\\
$GST\gets \emptyset$ \Comment{Initialize an empty tree}\\

\For{$T_i \in T$}
{
Identify a respective Latent Concept $L_i$ for each $T_i$\\
    $SteinerTree_i \gets$ \textsc{SteinerTree}$(G, L_i)$ \Comment{Calculate Steiner trees for vertices in the group $L_i$}\\
    $GST \gets GST \cup SteinerTree_i$ \Comment{Union with the resultant tree}\\
}
\For{$L_i \in L$}{
    \Comment{Repeat until the group is empty}\\
    Find the vertex $v$ in $Li$ with maximum number of edges in $GST$\\
    $T_v \gets$ \textsc{SteinerTree}$(G, v)$\\
    $GST \gets GST \cup T_v$ \Comment{Merging with the main tree}\\
    $L \gets L - \{L_i\}$ \Comment{Remove processed vertex from the group}\\
}

 $GST=(C^{*}, R^{*}, W^{*})$ where $R^{*}\subseteq R, C^{*} \subseteq C$  with $Min(\sum_{(i,j)\in R^{*}} w_{ij})$

\(Relevant\_Concepts=\)Nodes in \(GST\)\\
\textbf{return} $Relevant\_Concepts$ \Comment{Return the relevant concepts}
}
\end{algorithm}




\par In the proposed algorithm~\ref{alg2}, each Anchor Concept is extended starting to obtain the semantic group. Trees are developed iteratively by linking terminal nodes with the least-cost edges from each group. When common vertices are detected, trees are merged continuously. When a tree covers all Terminal nodes, the process is terminated. 
\par The optimum tree may be a structure with  Terminal nodes only. In such a case, the proposed algorithm will not generate a tree from semantic groups based on the GST. The number and distribution of terminal nodes in each group  influence the complexity of finding a solution to the Group Steiner Tree problem\cite{ApproxGST}.
\subsection{Complexity Analysis}
\par The most straightforward brute-force approach to solving the Steiner Tree problem is to enumerate all possible subsets of vertices to find the minimum-cost tree. This approach has an exponential time complexity of $O(2^n \cdot n^2)$, where $n$ is the number of vertices in the graph~\cite{complexity1,complexity2}. Enumerating all subsets of vertices has a complexity of $O(2^n)$, and for each subset, finding a minimum spanning tree has a complexity of $O(n^2)$.
\par Let's consider,  the Group Steiner Tree algorithm operates on a graph with $n$ vertices and $m$ edges, along with a group $K$ containing $k$ vertices. To solve the problem, the algorithm calculates Steiner trees for each vertex in the group, connecting them to the rest of the vertices in the graph. The time complexity of this step relies on the chosen algorithm, typically around $O(2^k \cdot n^2)$. The algorithm then merges these Steiner trees with a main tree, taking $O(n^2)$ time. Processed vertices are subsequently removed from the group, which can be done in $O(k)$ time. These steps are repeated until the group is empty, with the loop having an overall time complexity of $O(k \cdot (2^k \cdot n^2 + n^2 + k))$. Finally, the minimum group Steiner tree is returned in constant time, resulting in an overall time complexity of approximately $O(k \cdot 2^k \cdot n^2)$. It's worth noting that these complexities are estimates and may vary based on specific algorithms and heuristics employed.

\section{SemDR: Semantic-based Document Retrieval}
SemDR refers to a semantic-based document retrieval system that leverages the proposed algorithm to retrieve documents based on their semantic relationships. By incorporating domain information and employing semantic grouping through the Group Steiner Tree approach, SemDR aims to identify relevant concepts and improve the accuracy and comprehensiveness of document retrieval. It utilizes a knowledge graph constructed from domain-specific information and real-world data to enhance the retrieval process. SemDR is designed to overcome the limitations of traditional text-based search methods and provide more precise and meaningful search results for users. SemDR is typically composed of an Index generation process, Document Access, and Query processing. The indexing system generates an index structure by evaluating each document's text and finding key phrases and concepts. Documents are accessed via the index structure. The query processing system is responsible for the task of processing user queries over documents retrieved. To identify documents that best match the user's query, the system often includes a collection of algorithms like keyword matching, and semantic analysis. 
\begin{figure}[!h]
\centering
\includegraphics[width=88.9mm,height=50mm]{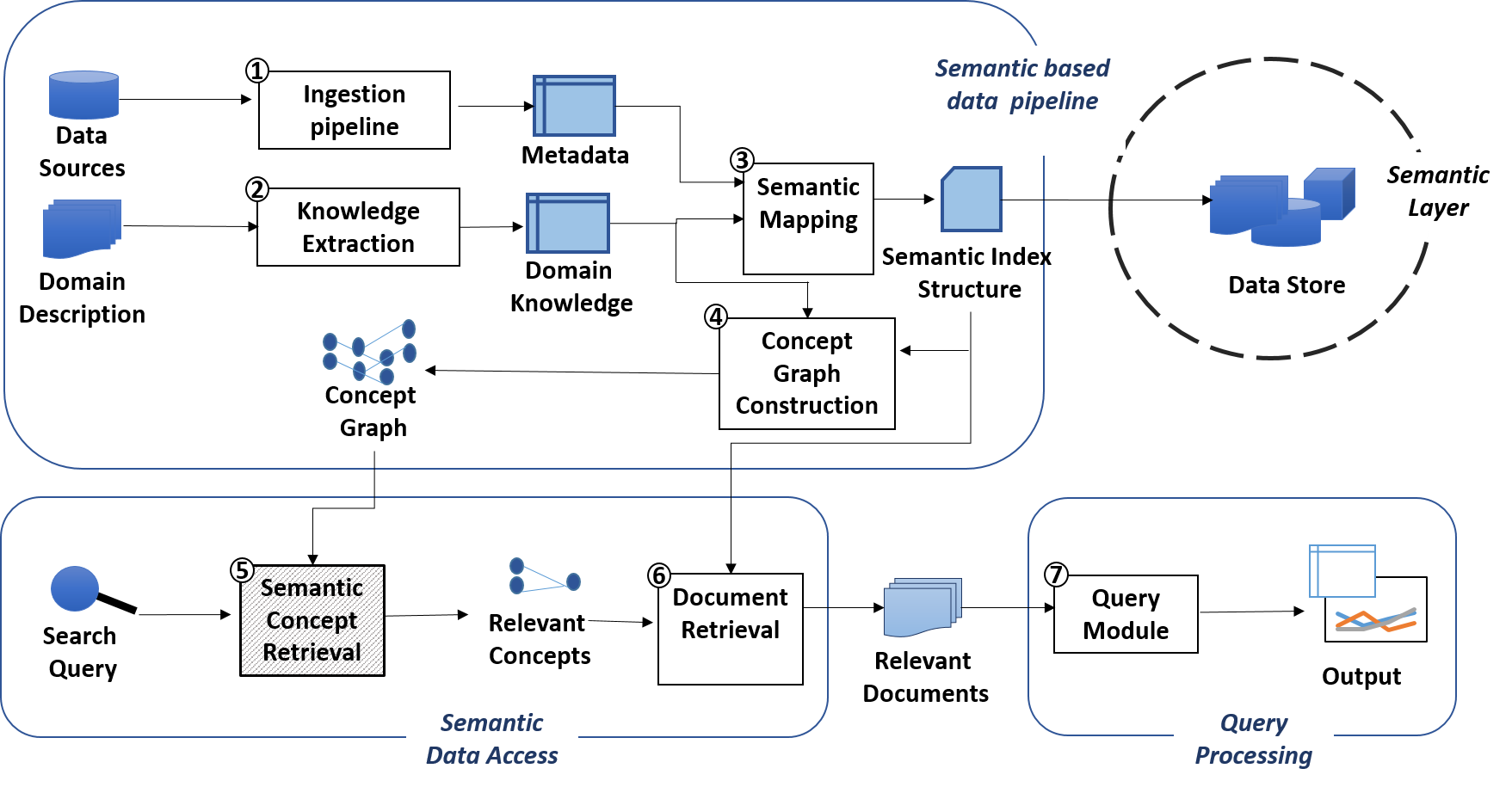}
\caption{ The architecture of SemDR system}
\label{f3}
\end{figure}
\par The proposed algorithm - `Semantic Concept Retrieval by GST' discussed in section III, is applied in a SemDR system. The key component of SemDR is the discovery of semantically relevant concepts and their utilization in document retrieval. Figure~\ref{f3} depicts an overview of the SemDR system. There are three modules in total:\\ 
1) The Semantic Pipeline is concerned with providing the framework for semantic data retrieval. \\
2) Semantic concept retrieval; this module is based on the algorithm we proposed. The module's finding of relevant concepts aids in the retrieval of related documents. \\
3) The query module enables the user to run structured linked queries over the obtained documents.
\subsection{Architecture overview} 
Semantic-based data pipeline module is responsible for generating index structure over heterogeneous data. The semantic index structure generates a semantic layer over data sources. All document retrieval requests are processed via index structure.\\
\textit{Data sources:} Table~\ref{t1} highlights the data sources contributing to the SemDR system. Two types of data sources are considered for implementation- structured and unstructured. The primary goal of the SemDR system is to pertain data sources in their original format while facilitating the retrieval mechanism. We extract the metadata about the document context to enable heterogeneous data for processing. The metadata captures the contextual characteristics of the document. In structured documents, it comprises attributes, domain, range, and description; unstructured documents primarily include frequent words and descriptions. The metadata acts as a logical representation of documents, which is used by the semantic mapping process to index the documents. 
\begin{table}[!ht]
  \centering
  \caption{Description of data sources}
  \begin{tabular}{|p{0.049\linewidth}|p{0.4\linewidth}|p{0.09\linewidth}|p{0.07\linewidth}|p{0.06\linewidth}|}
  \hline
    \textbf{Data } & \textbf{Description} & \textbf{Type} & \textbf{Files} & \textbf{Size 
    \tiny{(0.8GB)}}\\\hline
    DS1 & Districtwise fertilizer and crop production data, Karnataka. & Struct -ured & SQL & 6\% \\ \hline
    DS2 & Crop yield and seasonal crop production for each taluk, village and district in Karnataka & Struct -ured & CSV & 59\% \\ \hline
    DS3 & Agriculture domain data for taluks and districts in Karnataka. & Struct -ured & XLS, XLSX & 11\% \\ \hline
    DS4 & The files cover seasonal crop production for the years 2016 and 2017 in Karnataka. & Unstru -ctured & PDF & 6\% \\ \hline
    DS5 & The files explain about Data Project Specifications of land Utilisation Analysis & Unstru -ctured & DOC & 1\% \\ \hline
    DS6 & The image dataset covers information about millets, seasonal crops, and year-wise production in Karnataka & Unstru -ctured & JPEG, PNG & 17\% \\ \hline
  \end{tabular}
  \label{t1}
\end{table}
\\\textit{Domain Knowledge extraction:}
Semantic Concept Graph $G$ constitutes the foundation of our proposed algorithm and the SemDR system. The $G$ is built over the data gathered from domain knowledge. The subject matter expert compiles his years of expertise and insight into a document, report, or article. These artifacts aid in the development of domain knowledge. Government documents, survey information, performance reviews, and proposals are examples of artifacts that add to domain knowledge. $SPO$ triples representation is the most widely used representation to relate raw data, domain knowledge and semantics. $SPO$ triples are composed of three elements: The subject $S$ denotes the context. Predicate $P$ defines the relationship or attribute that characterizes the subject to the object $O$. $SPO$ triples are used to represent data in a machine-readable manner, giving systems the ability to deduce the relationships between entities. RDF/XML, Turtle, and JSON-LD are a few file types that can hold SPO triples. These file types can transfer data between systems or store data, making it simple for individuals to use. Additionally, $SPO$ triples can be stored in CSV (Comma Separated Values) file form. Each $SPO$ triple is displayed in this format as a row in the file, with commas separating the subject, predicate, and object values. 
\par The SemDR system accepts CSV and OWL formats to leverage the domain knowledge. The domain knowledge includes information for each entity, such as words that describe the entity, linked entities, and relationships. SemDR captures domain knowledge in $SPO$ format using CSV or OWL formats. SemDR treats subjects and objects as concepts and the description, superclass and subclass relationship among concepts as predicates. The information captured is referred to as concept information is further used for semantic mapping purposes.\\
\textit{Semantic mapping:} The semantic mapping stage is critical in constructing the semantic index structure. The index structure allows access to data while also contributing to creating a concept graph. The primary goal of semantic mapping is to connect documents to concepts. The semantic mapper considers two inputs, the first input is concept information and another input is document metadata. We can relate texts to concepts using the semantic similarity function. However, the goal here is not just to return all relevant documents but also to allow users to discover relevant concepts linked to their search query. The SemDR system proposes the clustering approach to group together documents that share common characteristics or themes, enabling us to identify new concepts and patterns within a collection of documents. 
\begin{figure}[!h]
  \begin{center}
   \includegraphics[width=70mm,height=30mm]{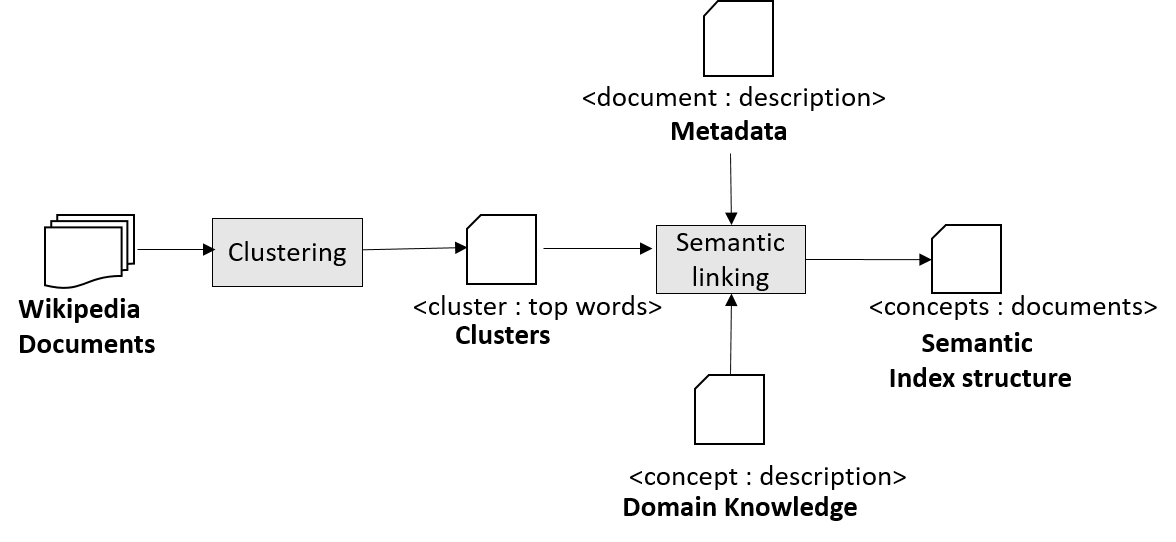}
   \caption{Semantic Mapping Process
   \label{f4}}
   \end{center}
  \end{figure}
\\\textit{Clustering:} In document retrieval, K-means clustering is used to group documents with similar content or features, streamlining the organization and retrieval of pertinent information. This approach facilitates the identification of document clusters sharing common themes or topics, enhancing the effectiveness of information retrieval processes.
\par If ingestion is considered to be a one-time process; grouping ingested documents and then selecting top words in each cluster to map them to concepts using a semantic similarity might be an effective option. However, given the dynamic nature of ingestion, the system should continue to accept new documents, which would result in changes to the clusters and index structure. A static generic cluster would be useful in overcoming this difficulty. We used Wikipedia data dump and basic k-means clustering to identify generic clusters in this implementation. We also found the top terms that define each cluster. 
\par The key observation here is that any given corpus can be utilized to discover generic clusters. However, the specific approach to clustering, the decision on the number of clusters, and the selection of the top words representing a cluster may vary depending on the specific domain of interest.
\par The Figure~\ref{f4} depicts the semantic mapping block in more detail. The semantic block is given domain knowledge containing concepts and their descriptions and metadata defining documents. The static clusters serve as the anchor element. Documents are mapped to clusters, and concepts are mapped to clusters, using a semantic similarity function that generates concept-to-document mapping as a semantic index structure.
\subsection{Semantic Graph Construction:}
 This block is responsible for building the concept graph. Section II has elaborated on the intuition involved in the process of graph construction. The domain knowledge is used to create a basic concept graph in which each concept is represented as a node, and the predicate defines the edge. The graph generated is a bidirectional unweighted graph. In this implementation, concept graph includes 327 concepts, and almost 86,000 edges. To assign weights to the graph, the system uses a generated index structure. Using the concept\_domain function, we compute a score for each concept. For each concept c in domain knowledge, the function $concept\_domain$ is defined as 
\begin{equation}
  \label{conc}
  concept\_domain(c)=CD_c=Semantic\_Index[c]
\end{equation}
\par The function value is represented by the set of documents associated with each concept. We define the function relation\_score to determine the score for every edge using inverse Jaccard similarity or the Jaccard index. 
It is a measure of how distinct the two groups of elements are. It represents the value of elements that differ between the two sets in relation to the total number of distinctive objects in both groups. 
The Jaccard similarity coefficient can be defined as, 
\begin{equation}
\label{Jac}
  J(CD_{C_1},CD_{C_2})=\frac{|CD_{C_1} \cap CD_{C_2}|}{ |CD_{C_1} \cup CD_{C_2}|}
\end{equation}
\
where \(C_1\),\(C_2\) are concepts and \(CD_{c_1}\,CD_{C_2}\) holds the set of documents associated with respective concepts defined by $concept\_domain$ function. 
The primary goal of assigning weights to relationships is to select concepts with the highest number of documents linked with them while deciding on the Group Steiner Tree in algorithm~\ref{alg2} described in Section III. In this application of document retrieval, the major focus is on identifying the nodes having a maximum number of associated documents. To achieve this over shortest path finding algorithms, We use the Inverse Jaccard similarity coefficient to define the \textit{relation\_score}. The inverse score assures the selection of concepts with a greater edge weight. The Inverse Jaccard similarity coefficient is defined as 1 minus the Jaccard similarity coefficient. 
The function \textit{relation\_score} can be defined mathematically as follows:
\begin{equation}
  \label{relsc}
  relation\_score(C_1,C_2) = 1 - J(CD_{C_1},CD_{C_2})
\end{equation}
\par The Semantic Concept Retrieval block receives the weighted semantic concept graph and applies the proposed algorithm in order to obtain the relevant concepts. The relevant concepts are forwarded to the data access block to retrieve the documents using a semantic index structure. The semantic closeness of the retrieved documents to the search input is used to rank the documents. 
\subsection{Query Module:}
Given that the data sources are heterogeneous, the retrieved and ranked documents may be in different file formats. In this study, we restrict ourselves to solely querying structured documents. The structured documents include CSV, XLS, XLSX, and SQL data files. These structured documents are semantically connected for querying. Based on feasibility, the documents that have been retrieved are merged. The merged documents are then considered for the join operation. The equivalent of a relational JOIN is performed between data files only if there is any link between them that is equivalent to a PRIMARY KEY and FOREIGN KEY relationship. The column name, attribute range, attribute domain values, and semantic similarity are all considered to determine the link condition. If attributes match the factors mentioned above, then the documents are considered for join queries.
The result is shown in both its visual representation and tabular form.
\section{Results and Discussion}
The objective of this section is to provide an overview of the conducted experiments, including details on the utilized datasets, and evaluation metrics. Furthermore, the significance of benchmarking is emphasized, as it enables an objective comparison with baseline systems. This section is organized around the following key points:
\textit{A. Experimental Setup, B.Baseline Systems, C.Evaluation Metrics, D.Result and Observations.}
\subsection{Experimental Setup}
We compare the results of the proposed system with the outcomes of the baseline systems. We are considering a keyword-based retrieval system and a semantic-based retrieval system as a baseline. Our experimental evaluation is limited to Lucene, ElasticSearch, and Doc2Vec tools. 
The reference solution is built considering the application of the proposed research.  The domain experts have asked to select files from the search corpus for all the search strings. These results are considered reference solutions to compare with baseline systems and the proposed approach.
\subsubsection{Search Query}
The search query represents the user's search interest in retrieving documents. We have collected the search queries from domain experts to test the recommended system considering the combination of search interests including geographical location, time parameter and search strings. The search queries are gathered using Google Forms. To accommodate search strings based on standard test cases relevant to the domain, we refer to use-cases provided by government use-case analysis\footnote{NDAP-https://ndap.niti.gov.in/}. Table~\ref{t2} contains information about the search queries. The search strings can be a combination of domain interest, time period of interest, and geographical location (state/district/village). 
 \par To understand the performance of the proposed system in more detail, we group search queries as follows: keywords appearing in the dataset called Direct search words, keywords with geographic location, keywords with the time period, and strings that are not appearing in the dataset but semantically relevant to  the dataset called as Indirect search words. All of these types contribute to a total of 170 search strings.
 \begin{table}[!ht]
    \centering
    \caption{Description of Search Query}
    \begin{tabular}{|p{0.064\linewidth}|p{0.43\linewidth}|p{0.12\linewidth}|p{0.036\linewidth}|p{0.099\linewidth}|}\hline
        \textbf{Query set} & \textbf{Description} & \textbf{Type of queries} & \textbf{Size} & \textbf{Users}  \\ \hline
        QS1 & The queries in this set are mainly descriptive. The query words are majorly part of the search space. & Direct search words & 25 & Domain Users  \\ \hline
        QS2 & The queries are tagged with geo entities like Taluk, Village, or District in Karnataka. & Direct with Geo-tags & 11 & Domain Users \\ \hline
        QS3 & The queries are combinations of description, geo-tagging, and time parameters as year mentioned in yyyy format. & Direct with time tags & 24 & Domain Users  \\ \hline
        QS4 & The description used in this set of queries may not be directly present in search space but semantically highly relevant to the domain. & Indirect  query & 6 & Domain Users  \\ \hline
        QS5 & These sets of queries are generated from use-cases from domain experts.  & Direct-Indirect query & 104 & Domain Experts\\ \hline
    \end{tabular}
    \label{t2}
\end{table}
\subsubsection{Reference Solution }
We consider public data for implementation, as indicated in Table~\ref{t1}. The information is gathered from multiple government departments. Our search query corpus and concept graph consider domain experts and domain users as sources of input because they are the primary sources of authentic information. In SemDR, 
the reference solution is built over a query set as discussed in Table \ref{t2}. The domain experts selected the files of interest by considering metadata and content. The search query and the set of relevant files selected by the domain expert contribute to the reference solution. 
\subsection{Baseline Systems} 
We opted for the following three baseline systems for the performance analysis of SemDR: \\
\textit{Lucene} is an open-source search library written in Java that provides a collection of low-level APIs for indexing and searching data~\cite{Lucene1}. Lucene allows users to index documents by analysing the content and constructing a keyword-based search index~\cite{Lucenefordr}.\\
\textit{Elasticsearch} leverages Lucene as its fundamental search engine with additional capabilities for distributed search, real-time search, analytics, and queries in domain-specific languages~\cite{es}.\\
\textit{Doc2Vec} is a natural language processing technique that takes the Word2Vec algorithm and applies it to entire documents. Document embeddings are dense vector representations of whole documents in the same high-dimensional space that Doc2Vec generates. The vector space approach can also be used to find relevant materials~\cite{doc2vec}.
\subsection{Evaluation Metric} 
To validate the results, we analyze conventional parameters such as precision, accuracy, recall, F1-score, and error score against each search string. To compute all of the above parameters, the manual selection results by the experts are used as actual values, while the baseline or SemDR system results are used as predicted values.

The confusion matrix is built on the calculation of the following values, \\
\textit{True Positive (TN)} indicates the count of documents retrieved by the manual selection process and other baseline systems. \\
\textit{False Positive (FP)} indicates the number of documents retrieved by the baseline system but not by the manual selection process. \\
\textit{False Negative (FN)} indicates the number of documents retrieved by the manual selection process but not by the baseline system. \\
\textit{True Negative (TN)} indicates the count of documents neither retrieved by the baseline system nor by the manual selection process. \\
\textit{Error Type.} When a document retrieval system retrieves irrelevant documents that shouldn't have been retrieved, the error is known as a \textit{type 1 error}. In other words, it is a false positive error when a search result includes a document that is not related to the user's search.
The \textit{type 2 error} occurs in document retrieval when the system fails to retrieve a relevant document that should have been retrieved. It is a false negative error in which a document that is related to the user's query is not included in the search results.\\
The performance metrics derived from the confusion matrix are:\\
\textit{Accuracy:} In document retrieval systems, accuracy measures the percentage of retrieved documents relevant to the query. More formally, accuracy is the ratio of correctly retrieved documents to the total number of retrieved documents. 
\(Accuracy= (TP + TN) / (TP + TN + FP + FN)\)\\
\textit{Precision:} Precision measures the proportion of retrieved documents relevant to the query. It is the ratio of correctly retrieved documents to the total number of retrieved documents. In other words, precision measures how precise the system is in returning only relevant documents.
\(Precision= TP / (TP + FP)\)\\
\textit{Recall:} Recall measures the proportion of relevant documents retrieved by the system. It is the ratio of correctly retrieved documents to the total number of relevant documents. Recall measures how effective the system is in retrieving all relevant documents. 
\(Recall/Sensitivity= TP / (TP + FN)\)\\
 \textit{F1 score:} F1 score is a measure that combines precision and recall to provide a single metric that balances both measures. It is the harmonic mean of precision and recall. 
\(F1-score= 2 * ((precision * recall) / (precision + recall))\)\\

\par We consider the above parameters to understand the overall performance of the system where accuracy measures the overall effectiveness of the system, precision measures the proportion of relevant documents returned, recall measures the proportion of relevant documents retrieved, and the F1 score combines precision and recall to provide a balanced measure of the system's effectiveness.
\par Within the context of evaluation, we have undertaken the conversion of metrics such as precision, recall, accuracy, and F1-score into a percentage scale. This conversion serves to facilitate the interpretation and comparative analysis of performance for various experiments across baseline systems.
\subsection{Result and Observations}
\begin{figure}
    \begin{center}
      \includegraphics[width=85.9mm,height=40mm]{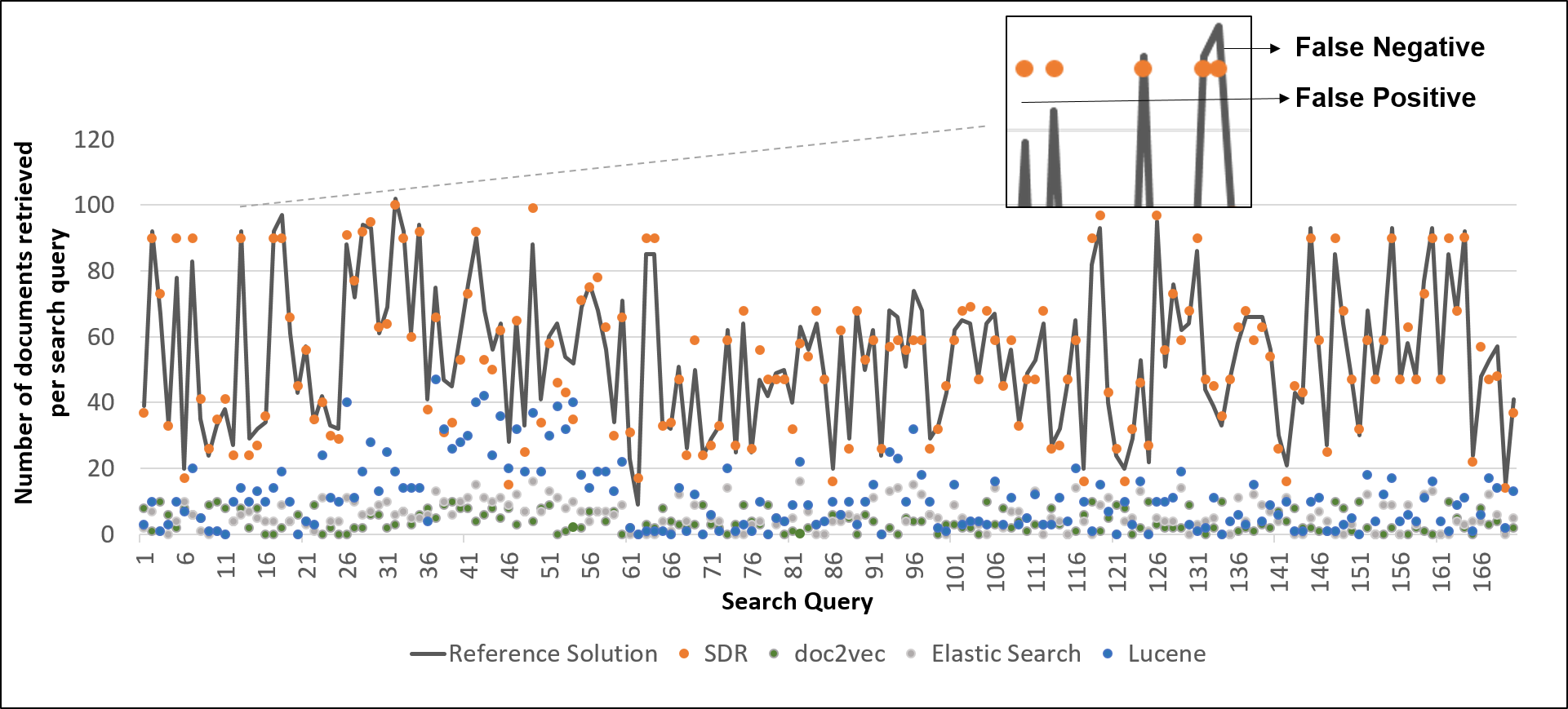}
      \caption{Document Retrieval Trend
      \label{f6}}
      \end{center}
    \end{figure}

Figure~\ref{f6} compares the baseline systems to the SemDR system. The trend line depicts a reference solution against 170 search keywords.  The graph indicates that the red/orange dots denoting the results of the SemDR system follow the trend line, whilst other systems retrieve fewer documents being clustered near the X-axis.
\subsubsection{Evaluation of SemDR using conventional performance parameters}  
The graph in Figure~\ref{f10} indicates the performance evaluation of four different systems (SemDR, Elastic Search, Lucene, and Doc2Vec) based on four different evaluation metrics (precision, recall, accuracy, and F1-score). To ensure a standardized comparison scale, all values are expressed as percentages. The performance of SemDR is quite good, with precision and recall at 90\% and 88\%, respectively, accuracy at 82\%, and an F1-score of 89\%. This indicates that SemDR retrieves a high percentage of relevant documents with a good balance of precision and recall. On the other hand, Elastic-Search, Lucene, and Doc2Vec have lower precision, recall, accuracy, and F1-score values. Elastic-Search has the highest precision among these three systems but with a low recall, indicating that it retrieves only a small number of relevant documents. Lucene and Doc2Vec have lower precision and recall values than Elastic-Search, and they also have lower accuracy and F1-score values, indicating poor overall performance. 


\subsubsection{Error Calculation:} 
In general, it is ideal for minimizing both type 1 and type 2 errors. But some error types could be more severe than others, depending on the application. It is challenging to guarantee the accuracy of the knowledge used in implementing SemDR because it is based on domain knowledge provided by subject matter experts. Additionally, the actual outcomes used for evaluation are interpreted manually, which introduces drift into the results. Therefore, it is essential to comprehend the result with documents neither retrieved by the baseline nor by the proposed system.  In Figure~\ref{f6}, the error is represented by the zoomed section. To understand the trend of not retrieved documents,  \textit{type 2 error} or \textit{false negative error} is considered. With reference to Table~\ref{table:error_comparison}, the error value measures how often the system produces incorrect results, expressed as a percentage of the total number of queries processed. The mean of Type-2 (False Negative) Error is calculated as \[\sum_{\text{search queries}} \frac{(\text{Reference solution - retrieved documents})}{\text{reference solution}}\] \par It is observed that Lucene has an error value of 80.2\%, ElasticSearch has an error value of 88.1\%, Doc2Vec has a value of 96\%, and SemDR has a value of 11.9\%. The SemDR system has the lowest error rate, which means it has the highest accuracy among the four systems.\\
\begin{table}[h]
\centering
\caption{Comparison of Type-2 Error}
\begin{tabular}{|c|c|c|c|c|}
\hline
\textbf{System }& SemDR & Elastic Search & Lucene & Doc2Vec \\
\hline
\textbf{Type-2 Error} & 11.9 & 88.1 & 80.2 & 96 \\
\hline
\end{tabular}

\label{table:error_comparison}
\end{table}
\subsubsection{Comparison with baseline systems} 
\begin{figure}[!h]
\centering
\includegraphics[width=88.9mm,height=80mm]{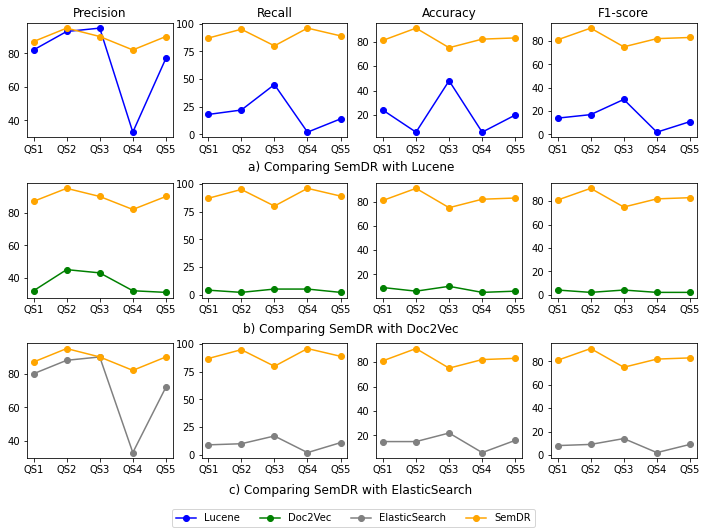}
\caption{Comparing SemDR with other baseline systems}
\label{f10}
\end{figure}

Figure~\ref{f10} illustrates the True Positive value over different baseline systems with respect to different search strings with location and time tags. It is observed that the overall performance of the systems varied across query types. SemDR had the highest overall performance, retrieving relevant results for the queries across all query types. Lucene and Elastic Search also performed well, particularly for Direct Query searches with additional location and time tags (QS2 and QS3). Doc2Vec had relatively low performance across all query types. 

Figure~\ref{f10} illustrates the performance parameters to comprehend how different systems perform with respect to different search strings with location and time tags. The orange line indicates the trend of the SemDR system. All the graphs capture various parameters over a query set QS1, QS2, QS3, QS4 and QS5. It is observed that the orange line trends achieve better values over all the parameters. Referring to the graph in Figure~\ref{f10}, we can see that SemDR performs the best overall, with the highest scores in precision, recall, accuracy, and F1-score across all five types of queries.


\subsubsection{Evaluation with ranked retrieval}
\par Document retrieval is an ideal application for managing systems with massive amounts of data. The ranking mechanism would help users focus on important documents based on search interest. The retrieved documents are organized using ranking algorithms based on their relevance to the user's input, allowing the user to query the most pertinent data rapidly. To define the ranking, the proposed method uses the cumulative score of the semantic similarity function across the search string and document metadata. 
All the considered baseline systems generate ranked retrieval of the documents. As the reference system is based on manual intervention, the rank of the document is a subjective value and would vary for each domain expert. Hence, it is critical to calculate the overall ranking error. 
\par In our experimental evaluation, we try to choose top-k ranked documents and check if they exist in the result set of the reference system. This validates the baseline systems over ranked retrieval.  We analyzed the top 3, 5, 7, and 10 retrieved documents for comparison. To evaluate the ranking of documents over 170 queries, the mean of True Positive values in percentage is calculated as, \[\sum_{\substack{\text{search queries}}} \left| \frac{\text{Reference solution} - \text{retrieved documents}}{\text{top retrieved documents}} \right|
\] \par Table~\ref{f11} indicates that as the number of considered documents increases from 3 to 10, the percentage of relevant documents retrieved by other baseline systems decreases except the SemDR system signifying the proposed system retrieves more relevant documents.
\begin{table}[h]
\centering
\caption{Analysis of ranked retrieval over baseline systems}
\begin{tabular}{|c|c|c|c|c|}
\hline
 & SDR & Elastic Search & Lucene & Doc2Vec \\
\hline
Top-3 & 75 & 75 & 80 & 30 \\
Top-5 & 82 & 70 & 71 & 24 \\
Top-7 & 86 & 61 & 65 & 20 \\
Top-10 & 89 & 51 & 61 & 16 \\
\hline
\end{tabular}

\label{f11}
\end{table}

    
\subsubsection{Additional ontological Constructs}
To adapt the proposed algorithm for document retrieval over government data, we observe that search strings include various geographic locations and time references to refine the search interest. It is also observed that the search queries mention Village, District, or Taluk to refine their search interest. The logical representation of the geography of India includes a hierarchy of (Country-State-District-Taluk-Village)\footnote{The generic geographic hierarchy is Country-State-County-Village, here Taluk can be considered equivalent to County}. We also capture information about neighboring Villages, Talukas, and Districts. To support geographically tagged search, we captured geographic information in the form of the ontology using shape files and constructed separate index structures. We preferred to keep the domain ontology and geographic ontology separate for the following reasons, 
\begin{itemize}
    \item The considered Agricultural domain knowledge is generic. Still, the data used for ingestion and search queries are geographically tagged and relevant to Karnataka.
    \item The generic domain ontology prefers identifying semantically grouped concepts, but the geographic ontology holds facts and hierarchical structure.
    \item To identify the alternate concepts in the geographic ontology, we prefer traversing to the neighboring node or the parent node and not a semantically relevant node.
\end{itemize}

\section{Related Literature}
The choice of a specific retrieval mechanism is influenced by multiple factors, including document characteristics, user requirements, and available resources for system development. Depending on the search corpus, cross-lingual, text-based, or image-based systems can be employed. Traditional methods utilize diverse techniques like natural language processing, machine learning, statistical models, neural networks, and semantic-based models to extract pertinent information.\\
\textbf{Approaches for Retrieval:} Traditional keyword-based search relies heavily on exact keyword matches~\cite{k0}. Conversely, Google's BERT model for NLP considers contextual meaning, enhancing keyword-based search~\cite{k1,k2}. Research~\cite{k3} introduces a BERT-based model with adaptive weighting for query terms, outperforming traditional methods treating all terms equally. For efficient keyword-based retrieval on large-scale text data~\cite{k4}, a two-stage approach is proposed, initially retrieving candidates with an inverted index and then ranking using machine learning. However, keyword-based models are prone to synonym and homonym challenges, potentially missing crucial data.
Concept-based models aim to understand query concepts beyond textual context, offering more accurate and comprehensive results. A novel neural network-based approach~\cite{c1} combines traditional vector space models with neural embeddings. A hierarchical network-based approach~\cite{c2} captures global and local document information. Another hybrid approach~\cite{c3} fuses semantic knowledge graphs with vector space models. Nonetheless, concept-based models may lack coverage for general concepts as they're often trained on specific datasets.
Semantic models capture rich semantic connections, including synonyms and antonyms, enabling thorough and precise results, especially for complex searches~\cite{s1,s2}. Research~\cite{base} proposes a semantic-based retrieval method using WordNet and ontology. Another work~\cite{s3} employs knowledge graphs for entity relationships. The Multi-Level Semantic Matching approach~\cite{s4} combines various features for similarity computation, often employing multiple models for improved accuracy~\cite{h1,h2,h3,h4}.\\
\textbf{Retrieval Models: }Various models are employed in document retrieval, including:
Boolean Model: It uses Boolean logic to match documents with queries but lacks ranking and term weighting, impacting retrieval effectiveness~\cite{bool1,bool2,bool3,bool4}. Vector Space Model (VSM) assigns terms unique dimensions and computes document-query similarity based on vector angles~\cite{vec1,vec2}. However, it struggles with semantic connections between terms~\cite{vec3}. Probabilistic Models employ probability theory to estimate document relevance based on term frequencies and document characteristics~\cite{prob1,prob2}. NLP and Language Models are pre-trained language models that learn contextualized representations for queries and documents, enhancing retrieval~\cite{nlp1,nlp2,nlp3}. Neural models extend these capabilities~\cite{neu1,neu2,neu3}. Machine Learning Models utilize labelled data to learn patterns and optimize parameters for retrieval~\cite{ml1,ml2}. Knowledge Graph (KG) Model represents information as a graph with entities as nodes and relationships as edges, offering structured knowledge for retrieval~\cite{kwg1,kwg2}.\\
\textbf{KG for Retrieval:} Knowledge graphs enhance semantic document retrieval by capturing meaning, providing context, supporting machine learning models, and offering advanced search capabilities~\cite{kg3,st1,st,st3,gst1,gst2,find}.\\
\textbf{Existing Search Engines: }While Document Retrieval (DR) and Information Retrieval (IR) are distinct, "Search Engines" represent another related branch widely used in web applications~\cite{searchEngine}. Semantic search engines, like Semantic Search-Nordlys, leverage various technologies for entity recognition, knowledge base linking, and relevant entity retrieval~\cite{nordlys}. The SCORE platform creates a large knowledge base using semantics~\cite{score}. Schema.org facilitates ontology-driven Semantic Search~\cite{schemaorg}. Existing search engines, however, may struggle with complex queries and semantic nuances, impacting result accuracy.
\par The literature extensively covers retrieval approaches and models. However, the absence of domain-specific context in retrieval can lead to misleading outcomes.
\section{Conclusion}
The research work introduces SemDR, an innovative semantic-based document retrieval system designed for heterogeneous data sources. The retrieval process is motivated by domain-specific information, which is harnessed through the process of concept graphs and the semantic linking of concepts. The system optimizes retrieval by employing semantic-driven concept identification using the Group Steiner Tree approach. This technique aids in pinpointing pertinent concepts and utilizing them for document retrieval.

The query module of SemDR facilitates the integration and querying of structurally relevant documents with semantic coherence. Notably, the SemDR system exhibits remarkable performance compared to conventional keyword-based retrieval systems and existing semantic-based document retrieval systems, attaining an accuracy rate of 90\% and a precision score of 82\%. As part of future work, the aspiration is to evolve SemDR to adapt seamlessly across diverse domains.


\section*{Acknowledgments}
This work was supported by Karnataka Innovation \& Technology Society, Dept. of IT, BT and S\&T, Govt. of Karnataka, India, vide GO No. ITD 76 ADM 2017, Bengaluru; Dated 28.02.2018. The research team is also grateful to the Government of Karnataka, the IIIT Bangalore Center for Open Data Research (CODR),  Bengaluru, India, for their significant data and domain expertise collaboration.



 
%


\bibliographystyle{IEEEtran}

\bibliography{references}

\begin{thebibliography}{10}
\providecommand{\url}[1]{#1}
\csname url@samestyle\endcsname
\providecommand{\newblock}{\relax}
\providecommand{\bibinfo}[2]{#2}
\providecommand{\BIBentrySTDinterwordspacing}{\spaceskip=0pt\relax}
\providecommand{\BIBentryALTinterwordstretchfactor}{4}
\providecommand{\BIBentryALTinterwordspacing}{\spaceskip=\fontdimen2\font plus
\BIBentryALTinterwordstretchfactor\fontdimen3\font minus \fontdimen4\font\relax}
\providecommand{\BIBforeignlanguage}[2]{{%
\expandafter\ifx\csname l@#1\endcsname\relax
\typeout{** WARNING: IEEEtran.bst: No hyphenation pattern has been}%
\typeout{** loaded for the language `#1'. Using the pattern for}%
\typeout{** the default language instead.}%
\else
\language=\csname l@#1\endcsname
\fi
#2}}
\providecommand{\BIBdecl}{\relax}
\BIBdecl

\bibitem{i1}
H.~Chu, \emph{Information representation and retrieval in the digital age}, 2003, information Today, Inc., url: \url{https://books.google.co.in/books?id=Rzg6WagUrawC&printsec=frontcover&source=gbs_ge_summary_r&cad=0#v=onepage&q&f=false}.

\bibitem{k0}
M.~Mitra and B.~Chaudhuri, ``Information retrieval from documents: A survey,'' \emph{Information retrieval}, vol.~2, pp. 141--163, 2000, doi: \url{https://doi.org/10.1023/A:1009950525500}.

\bibitem{k1}
C.~J. Yili~Qian and Y.~Liu, ``Bert-based text keyword extraction,'' \emph{Journal of Physics: Conference Series}, 2021, doi:10.1088/1742-6596/1992/4/042077.

\bibitem{k2}
B.~Gündoğdu, B.~Yusuf, and M.~Saraçlar, ``Joint learning of distance metric and query model for posteriorgram-based keyword search,'' \emph{IEEE Journal of Selected Topics in Signal Processing}, vol.~11, no.~8, pp. 1318--1328, 2017, \url{doi:10.1109/JSTSP.2017.2762080}.

\bibitem{k3}
Z.~Liu, W.~He, Y.~Wang, and H.~Ji, ``Adaptive document retrieval with query term dependency modeling,'' \emph{arXiv}, 2021, url: \url{https://aclanthology.org/D18-1055.pdf}.

\bibitem{k4}
H.~Chen, J.~Yan, H.~Jin, Y.~Liu, and L.~M. Ni, ``Tss: Efficient term set search in large peer-to-peer textual collections,'' \emph{IEEE Transactions on Computers}, vol.~59, no.~7, pp. 969--980, 2010, \url{doi:10.1109/TC.2010.81}.

\bibitem{c1}
H.~Chen, K.~J. Lynch, K.~Basu, and T.~D. Ng, ``Generating, integrating, and activating thesauri for concept-based document retrieval,'' \emph{IEEE Expert}, vol.~8, no.~2, pp. 25--34, 1993, \url{doi:10.1109/64.207426}.

\bibitem{c2}
S.~M. Sadjadi, H.~Mashayekhi, and H.~Hassanpour, ``A semi-supervised framework for concept-based hierarchical document clustering,'' \emph{World Wide Web}, pp. 1--30, 2023, doi: \url{https://doi.org/10.1007/s11280-023-01209-4}.

\bibitem{c3}
Y.~Yoon, K.~Choi, G.~Kim, and D.~Shin, ``A hybrid knowledge-based approach to information retrieval,'' \emph{Microprocessing and Microprogramming}, vol.~35, no. 1-5, pp. 329--336, 1992, doi: \url{https://doi.org/10.1016/j.cie.2022.108940}.

\bibitem{s1}
V.~Sugumaran and V.~C. Storey, ``A semantic-based approach to component retrieval,'' vol.~34, no.~3, 2003, \url{doi:10.1145/937742.937745}.

\bibitem{s2}
A.~Sharma and S.~Kumar, ``Ontology-based semantic retrieval of documents using word2vec model,'' \emph{Data \& Knowledge Engineering}, vol. 144, p. 102110, 2023, doi: \url{https://doi.org/10.1016/j.datak.2022.102110}.

\bibitem{s3}
X.~Zhang, X.~Hou, X.~Chen, and T.~Zhuang, ``Ontology-based semantic retrieval for engineering domain knowledge,'' \emph{Neurocomputing}, vol. 116, pp. 382--391, 2013, doi: \url{https://doi.org/10.1016/j.neucom.2011.12.057}.

\bibitem{s4}
M.~Hamroun, S.~Lajmi, M.~Jallouli, and A.~Souid, ``Efficient text-based query based on multi-level and deep-semantic multimedia indexing and retrieval,'' \emph{Multimedia Tools and Applications}, pp. 1--40, 2023, doi: \url{https://doi.org/10.1007/s11042-023-17256-y}.

\bibitem{bool1}
G.~Salton, E.~A. Fox, and H.~Wu, ``Extended boolean information retrieval,'' \emph{Communications of the ACM}, vol.~26, no.~11, pp. 1022--1036, 1983, url: \url{https://dl.acm.org/doi/pdf/10.1145/182.358466}.

\bibitem{bool2}
J.~H. Lee, ``Properties of extended boolean models in information retrieval,'' in \emph{SIGIR’94: Proceedings of the Seventeenth Annual International ACM-SIGIR Conference on Research and Development in Information Retrieval, organised by Dublin City University}.\hskip 1em plus 0.5em minus 0.4em\relax Springer, 1994, pp. 182--190, doi: \url{https://doi.org/10.1007/978-1-4471-2099-5_19}.

\bibitem{vec1}
A.~Neelakantan, J.~Shankar, A.~Passos, and A.~McCallum, ``Efficient non-parametric estimation of multiple embeddings per word in vector space,'' \emph{Conference on Empirical Methods in Natural Language Processing (EMNLP)}, 2015, doi: https://doi.org/10.3115/v1/D14-1113.

\bibitem{vec2}
S.~Clark, ``Vector space models of lexical meaning,'' \emph{The Handbook of Contemporary semantic theory}, pp. 493--522, 2015, doi: \url{https://doi.org/10.1002/9781118882139.ch16}.

\bibitem{ml1}
N.~Ireson, F.~Ciravegna, M.~E. Califf, D.~Freitag, N.~Kushmerick, and A.~Lavelli, ``Evaluating machine learning for information extraction,'' in \emph{Proceedings of the 22nd international conference on Machine learning}, 2005, pp. 345--352, doi: \url{https://doi.org/10.1145/1102351.1102395}.

\bibitem{prob2}
C.~Zhai and J.~Lafferty, ``A study of smoothing methods for language models applied to ad hoc information retrieval,'' in \emph{ACM SIGIR Forum}, vol.~51, no.~2.\hskip 1em plus 0.5em minus 0.4em\relax ACM New York, NY, USA, 2017, pp. 268--276, doi: \url{https://doi.org/10.1145/3130348.3130377}.

\bibitem{cikm}
A.~Kulkarni, C.~Ramanathan, and V.~E. Venugopal, ``Cognitive retrieve: Empowering document retrieval with semantics and domain-specific knowledge graph,'' 2023.

\bibitem{p1}
S.~Ji, S.~Pan, E.~Cambria, P.~Marttinen, and S.~Y. Philip, ``A survey on knowledge graphs: Representation, acquisition, and applications,'' \emph{IEEE transactions on neural networks and learning systems}, vol.~33, no.~2, pp. 494--514, 2021, iEEE, url: \url{https://arxiv.org/pdf/2002.00388.pdf}.

\bibitem{wu}
Z.~Wu and M.~Palmer, ``Verb semantics and lexical selection,'' \emph{arXiv}, 1994, doi: \url{https://doi.org/10.3115/981732.981751}.

\bibitem{gst1}
A.~Abujabal, X.~Lu, S.~Pramanik, R.~S. Roy, G.~Weikum, and Y.~Wang, ``Answering complex questions by joining multi-document evidence with quasi knowledge graphs,'' in \emph{SIGIR 2019}, 2019, url: \url{https://arxiv.org/abs/1908.00469}.

\bibitem{ApproxGST}
G.~Even and G.~Kortsarz, ``An approximation algorithm for the group steiner problem,'' in \emph{Proceedings of the thirteenth annual ACM-SIAM symposium on Discrete algorithms}, 2002, pp. 49--58.

\bibitem{complexity1}
C.~H. Papadimitriou and K.~Steiglitz, \emph{Combinatorial optimization: algorithms and complexity}.\hskip 1em plus 0.5em minus 0.4em\relax Courier Corporation, 1998, url: \url{https://dl.acm.org/doi/book/10.5555/31027}.

\bibitem{complexity2}
V.~V. Vazirani, \emph{Approximation algorithms}, 2001, vol. Springer, 1, url: \url{https://link.springer.com/book/10.1007/978-3-662-04565-7}.

\bibitem{Lucene1}
A.~Bia{\l}ecki, R.~Muir, G.~Ingersoll, and L.~Imagination, ``Apache lucene 4,'' in \emph{SIGIR 2012 workshop on open source information retrieval}, 2012, p.~17, url: \url{https://www.researchgate.net/profile/Andrzej-Bialecki/publication/260282732_Apache_Lucene_4/links/5ede6a2545851516e65f1fa6/Apache-Lucene-4.pdf#page=22}.

\bibitem{Lucenefordr}
P.~Yang, H.~Fang, and J.~Lin, ``Anserini: Enabling the use of lucene for information retrieval research,'' in \emph{Proceedings of the 40th international ACM SIGIR conference on research and development in information retrieval}, 2017, pp. 1253--1256, doi: \url{https://doi.org/10.1145/3077136.3080721}.

\bibitem{es}
C.~Gormley and Z.~Tong, \emph{Elasticsearch: the definitive guide: a distributed real-time search and analytics engine}.\hskip 1em plus 0.5em minus 0.4em\relax " O'Reilly Media, Inc.", 2015, url: \url{https://books.google.co.in/books/about/Elasticsearch_The_Definitive_Guide.html?id=d19aBgAAQBAJ&redir_esc=y}.

\bibitem{doc2vec}
J.~H. Lau and T.~Baldwin, ``An empirical evaluation of doc2vec with practical insights into document embedding generation,'' \emph{arXiv}, 2016, doi: \url{https://doi.org/10.48550/arXiv.1607.05368}.

\bibitem{base}
A.~Kulkarni, P.~Bassin, N.~S. Parasa, V.~E. Venugopal, S.~Srinivasa, and C.~Ramanathan, ``Ontology augmented data lake system for policy support,'' in \emph{Big Data Analytics in Astronomy, Science, and Engineering: 10th International Conference on Big Data Analytics, BDA 2022, Aizu, Japan, December 5--7, 2022, Proceedings}, 2023, pp. 3--16, springer, doi: 10.1007/978-3-031-28350-5\_1.

\bibitem{h1}
D.~Roy, D.~Ganguly, S.~Bhatia, S.~Bedathur, and M.~Mitra, ``Using word embeddings for information retrieval: How collection and term normalization choices affect performance,'' 2018, \url{doi:10.1145/3269206.3269277}.

\bibitem{h2}
T.~M.~M. Swe, ``Concept based intelligent information retrieval within digital library,'' 2021, doi: \url{http://dx.doi.org/10.2139/ssrn.3787840}.

\bibitem{h3}
D.~W. Ensan~F., ``Ad hoc retrieval via entity linking and semantic similarity,'' in \emph{Springer, Knowl Inf Syst 58, 551–583}, 2019, pp. 1205--1208, doi: \url{https://doi.org/10.1007/s10115-018-1190-1}.

\bibitem{h4}
W.~Hojas-Mazo, A.~Sim{\'o}n-Cuevas, M.~de~la Iglesia~Campos, F.~P. Romero, and J.~A. Olivas, ``A concept-based text analysis approach using knowledge graph,'' in \emph{International Conference on Information Processing and Management of Uncertainty in Knowledge-Based Systems}.\hskip 1em plus 0.5em minus 0.4em\relax Springer, 2018, pp. 696--708, doi: \url{https://doi.org/10.1007/978-3-319-91476-3_57}.

\bibitem{bool3}
B.~Hj{\o}rland, ``Classical databases and knowledge organization: A case for boolean retrieval and human decision-making during searches,'' \emph{Journal of the Association for Information Science and Technology}, vol.~66, no.~8, pp. 1559--1575, 2015, doi: \url{https://doi.org/10.1002/asi.23250}.

\bibitem{bool4}
M.~B. Aliyu, ``Efficiency of boolean search strings for information retrieval,'' \emph{American Journal of Engineering Research}, vol.~6, no.~11, pp. 216--222, 2017, url: \url{https://www.ajer.org/papers/v6(11)/ZA0611216222.pdf}.

\bibitem{vec3}
J.~Camacho-Collados and M.~T. Pilehvar, ``From word to sense embeddings: A survey on vector representations of meaning,'' \emph{Journal of Artificial Intelligence Research}, vol.~63, pp. 743--788, 2018, doi: \url{https://doi.org/10.1613/jair.1.11259}.

\bibitem{prob1}
A.~Berger and J.~Lafferty, ``Information retrieval as statistical translation,'' in \emph{ACM SIGIR Forum}, vol.~51, no.~2.\hskip 1em plus 0.5em minus 0.4em\relax ACM New York, NY, USA, 2017, pp. 219--226, doi: \url{https://doi.org/10.1145/312624.312681}.

\bibitem{nlp1}
F.~Song and W.~B. Croft, ``A general language model for information retrieval,'' in \emph{Proceedings of the eighth international conference on Information and knowledge management}, 1999, pp. 316--321, url: \url{https://dl.acm.org/doi/pdf/10.1145/319950.320022}.

\bibitem{nlp2}
U.~Naseem, I.~Razzak, S.~K. Khan, and M.~Prasad, ``A comprehensive survey on word representation models: From classical to state-of-the-art word representation language models,'' \emph{Transactions on Asian and Low-Resource Language Information Processing}, vol.~20, no.~5, pp. 1--35, 2021, doi: \url{https://doi.org/10.1145/319950.320022}.

\bibitem{nlp3}
S.~Borgeaud, A.~Mensch, J.~Hoffmann, T.~Cai, E.~Rutherford, K.~Millican, G.~B. Van Den~Driessche, J.-B. Lespiau, B.~Damoc, A.~Clark \emph{et~al.}, ``Improving language models by retrieving from trillions of tokens,'' in \emph{International conference on machine learning}.\hskip 1em plus 0.5em minus 0.4em\relax PMLR, 2022, pp. 2206--2240, url: \url{https://proceedings.mlr.press/v162/borgeaud22a.html}.

\bibitem{neu1}
B.~Mitra, N.~Craswell \emph{et~al.}, ``An introduction to neural information retrieval,'' \emph{Foundations and Trends{\textregistered} in Information Retrieval}, vol.~13, no.~1, pp. 1--126, 2018, doi: \url{http://dx.doi.org/10.1561/1500000061}.

\bibitem{neu2}
B.~Mitra and N.~Craswell, ``Neural models for information retrieval,'' \emph{arXiv}, 2017, doi: \url{https://doi.org/10.48550/arXiv.1705.01509}.

\bibitem{neu3}
J.~Guo, Y.~Fan, L.~Pang, L.~Yang, Q.~Ai, H.~Zamani, C.~Wu, W.~B. Croft, and X.~Cheng, ``A deep look into neural ranking models for information retrieval,'' \emph{Information Processing \& Management}, vol.~57, no.~6, p. 102067, 2020, doi: \url{https://doi.org/10.1016/j.ipm.2019.102067}.

\bibitem{ml2}
R.~Nallapati, ``Discriminative models for information retrieval,'' in \emph{Proceedings of the 27th annual international ACM SIGIR conference on Research and development in information retrieval}, 2004, pp. 64--71, doi: \url{https://doi.org/10.1145/1008992.1009006}.

\bibitem{kwg1}
R.~Reinanda, E.~Meij, M.~de~Rijke \emph{et~al.}, ``Knowledge graphs: An information retrieval perspective,'' \emph{Foundations and Trends{\textregistered} in Information Retrieval}, vol.~14, no.~4, pp. 289--444, 2020, doi: \url{http://dx.doi.org/10.1561/1500000063}.

\bibitem{kwg2}
F.~Corcoglioniti, M.~Dragoni, M.~Rospocher, and A.~P. Aprosio, ``Knowledge extraction for information retrieval,'' in \emph{The Semantic Web. Latest Advances and New Domains: 13th International Conference, ESWC 2016, Heraklion, Crete, Greece, May 29--June 2, 2016, Proceedings 13}.\hskip 1em plus 0.5em minus 0.4em\relax Springer, 2016, pp. 317--333, doi: \url{https://doi.org/10.1007/978-3-319-34129-3_20}.

\bibitem{kg3}
Q.~Guo, F.~Zhuang, C.~Qin, H.~Zhu, X.~Xie, H.~Xiong, and Q.~He, ``A survey on knowledge graph-based recommender systems,'' \emph{IEEE Transactions on Knowledge and Data Engineering}, vol.~34, no.~8, pp. 3549--3568, 2020, url: \url{10.1109/TKDE.2020.3028705}.

\bibitem{st1}
H.~Chen, H.~Wu, J.~Li, X.~Wang, and L.~Zhang, ``Keyword-driven service recommendation via deep reinforced steiner tree search,'' \emph{IEEE Transactions on Industrial Informatics}, 2022, doi: \url{10.1109/TII.2022.3177411}.

\bibitem{st}
C.~Chen, J.~Cui, G.~Liu, J.~Wu, and L.~Wang, ``Survey and open problems in privacy preserving knowledge graph: Merging, query, representation, completion and applications,'' 2020, doi: \url{https://doi.org/10.48550/arXiv.2011.10180}.

\bibitem{st3}
M.~W. Przewo{\'z}niczek, K.~Walkowiak, A.~Sen, M.~Komarnicki, and P.~Lechowicz, ``Solving the steiner tree problem for knowledge graphs using link prediction,'' vol. 479.\hskip 1em plus 0.5em minus 0.4em\relax Elsevier, 2019, pp. 1--19, doi: \url{https://doi.org/10.1016/j.ins.2018.11.015}.

\bibitem{gst2}
Y.~Shi, G.~Cheng, T.-K. Tran, E.~Kharlamov, and Y.~Shen, ``Efficient computation of semantically cohesive subgraphs for keyword-based knowledge graph exploration.''\hskip 1em plus 0.5em minus 0.4em\relax New York, NY, USA: Association for Computing Machinery, 2021, doi: 10.1145/3442381.3449900.

\bibitem{find}
X.~Ren, N.~Sengupta, X.~Ren, J.~Wang, and O.~Cur{\'e}, ``Finding minimum connected subgraphs with ontology exploration on large rdf data,'' \emph{IEEE Transactions on Knowledge and Data Engineering}, 2022, \url{doi:10.1109/TKDE.2022.3225076}.

\bibitem{searchEngine}
D.~Sharma, R.~Shukla, A.~K. Giri, and S.~Kumar, ``A brief review on search engine optimization,'' in \emph{2019 9th international conference on cloud computing, data science \& engineering (confluence)}.\hskip 1em plus 0.5em minus 0.4em\relax IEEE, 2019, pp. 687--692, \url{doi:10.1109/CONFLUENCE.2019.8776976}.

\bibitem{nordlys}
F.~Hasibi, K.~Balog, D.~Garigliotti, and S.~Zhang, ``Nordlys: A toolkit for entity-oriented and semantic search,'' in \emph{Proceedings of the 40th International ACM SIGIR Conference on Research and Development in Information Retrieval}, 2017, pp. 1289--1292, doi: \url{https://doi.org/10.1145/3077136.3084149}.

\bibitem{score}
M.~Sheth, M.~Riggs, and T.~Patel, ``Utility of the mayo end-stage liver disease (meld) score in assessing prognosis of patients with alcoholic hepatitis,'' \emph{BMC gastroenterology}, vol.~2, no.~1, pp. 1--5, 2002, doi: \url{https://doi.org/10.1186/1471-230X-2-2}.

\bibitem{schemaorg}
R.~V. Guha, D.~Brickley, and S.~Macbeth, ``Schema. org: evolution of structured data on the web,'' \emph{Communications of the ACM}, vol.~59, no.~2, pp. 44--51, 2016, \url{doi:10.1145/2844544}.

\end{thebibliography}
\end{document}